\documentclass[revtex4-1,apl,iop,twocolappendix]{emulateapj}
\usepackage[totalwidth=480pt, totalheight=680pt]{geometry}
\usepackage{graphicx}
\usepackage{amsmath,natbib}
\usepackage{geometry,natbib,float}
\def\eq{\begin{equation}}
\def\en{\end{equation}}

\def\etal{{\it et al.}\thinspace}

\def\eg{{\it e.g.,}\thinspace}

\def\etal{{\it et al}\thinspace}

\def\apj{{\it Ap.J.}\thinspace}
\def\apjl{{\it Ap.J. Lett.}\thinspace}
\def\apjs{{\it Ap.J. Suppl.}\thinspace}

\def\aap{{\it A\&A}\thinspace}

\def\mnras{{\it MNRAS}\thinspace}
\def\P3hat{{\mathaccent 94 P}_3}
\def\nat{{\it Nature}\thinspace}

\begin{document}

\title{Toward an Empirical Theory of Pulsar Emission XII: Exploring the \\
		Physical Conditions in Millisecond Pulsar Emission Regions}
\shorttitle{Exploring the Physical Conditions in Millisecond Pulsar Emission Regions}
\shortauthors{Rankin, Archibald, Hessels, van Leeuwen, Mitra, Ransom, Stairs, van Straten \& Weisberg}

\author{Joanna M. Rankin}
\affil{Physics Department, University of Vermont, Burlington, VT 05405 USA; email: Joanna.Rankin@uvm.edu \\
	Anton Pannekoek Institute for Astronomy, University of Amsterdam, Science Park 904, 1098 XH Amsterdam }
\author{Anne Archibald, Jason Hessels, Joeri van Leeuwen}
\affil{Anton Pannekoek Institute for Astronomy, University of Amsterdam, Science Park 904, 1098 XH Amsterdam \\
	ASTRON, the Netherlands Institute for Radio Astronomy, Postbus 2, 7990 AA, Dwingeloo, The Netherlands}
\author{Dipanjan Mitra}
\affil{Physics Department, University of Vermont, Burlington, VT 05405 USA \\
	National Centre for Radio Astrophysics, Ganeshkhind, Pune 411 007 India \\
	$^3$Janusz Gil Institute of Astronomy, University of Zielona G\'ora, ul. Szafrana 2, 65-516 Zielona G\'ora, Poland}
\author{Scott Ransom}
\affil{National Radio Astronomy Observatory, Charlottesville, VA 29201}
\author{Ingrid Stairs} 
\affil{Physics Department, University of British Columbia, V6T 1Z4, BC, Canada}
\author{Willem van Straten}
\affil{Institute for Radio Astronomy \& Space Research, Auckland University of Technology, Auckland 1142, New Zealand}
\author{Joel M. Weisberg} 
\affil{Physics \& Astronomy Department, Carleton College, Northfield, MN 55057}

\begin{abstract}
The five-component profile of the 2.7-ms pulsar J0337+1715 appears to exhibit the best 
example to date of a core/double-cone emission-beam structure in a millisecond pulsar 
(MSP).  Moreover, three other MSPs, the Binary Pulsar B1913+16, B1953+29 and J1022+1001, 
seem to exhibit core/single-cone profiles.  These configurations are remarkable and 
important because it has not been clear whether MSPs and slow pulsars exhibit similar 
emission-beam configurations, given that they have considerably smaller magnetospheric 
sizes and magnetic field strengths.  MSPs thus provide an extreme context for studying pulsar 
radio emission.  Particle currents along the magnetic polar flux tube connect processes 
just above the polar cap through the radio-emission region to the light-cylinder and the 
external environment.  In slow pulsars radio-emission heights are typically about 500 km 
around where the magnetic field is nearly dipolar, and estimates of the physical conditions there 
point to radiation below the plasma frequency and emission from charged solitons by the 
curvature process.  We are able to estimate emission heights for the four MSPs and carry 
out a similar estimation of physical conditions in their much lower emission regions.  We 
find strong evidence that MSPs also radiate by curvature emission from charged solitons.  
\end{abstract}


\section{Introduction}
Millisecond pulsar (MSP) emission processes have remained enigmatic despite 
the very prominence of MSPs as indispensable tools of contemporary physics and 
astrophysics.  MSP magnetospheres are much more compact ($\sim$100 to 
3000 km), and their magnetic field strengths are typically 10$^3$-10$^4$ times 
weaker and probably more complex than those of normal pulsars, due to the 
period of accretion that is believed to recycle old pulsars into MSPs.  

Normal pulsars are found to exhibit many regularities of profile form and polarization 
that, together, suggest an overall beam geometry comprised by two distinct emission 
cones and a central core beam.  It would seem that these regularities occur in major 
part because the magnetic field, at the roughly 500-km height where radio emission 
occurs, is usually highly dipolar (Rankin, Melichidze \& Mitra 2017).  

The vast majority of MSPs show no such regularity.  Their polarized profile forms 
are often broad and complex and seem almost a drunken parody of the order that 
is exhibited by many slower pulsars [\eg Dai \etal\ (2015) and the references therein].  
The reasons for this disorder are not yet fully clear, but the smaller  magnetospheres 
of MSPs may well entail emission heights where the magnetic fields are dominated 
by quadrupoles and higher terms.  Also, aberration/retardation (hereafter A/R) becomes 
ever more important for faster pulsars.  In this context, it is arresting to encounter a few 
MSPs that appear to exhibit the regular profile forms and perhaps polarization of 
normal pulsars.  If so, even a few such MSPs provide an opportunity to work out their 
quantitative emission geometry and assess whether their radiative processes are 
similar to those of normal pulsars.  

\begin{figure}
\begin{center}
\mbox{\includegraphics[width=80mm,angle=0.]{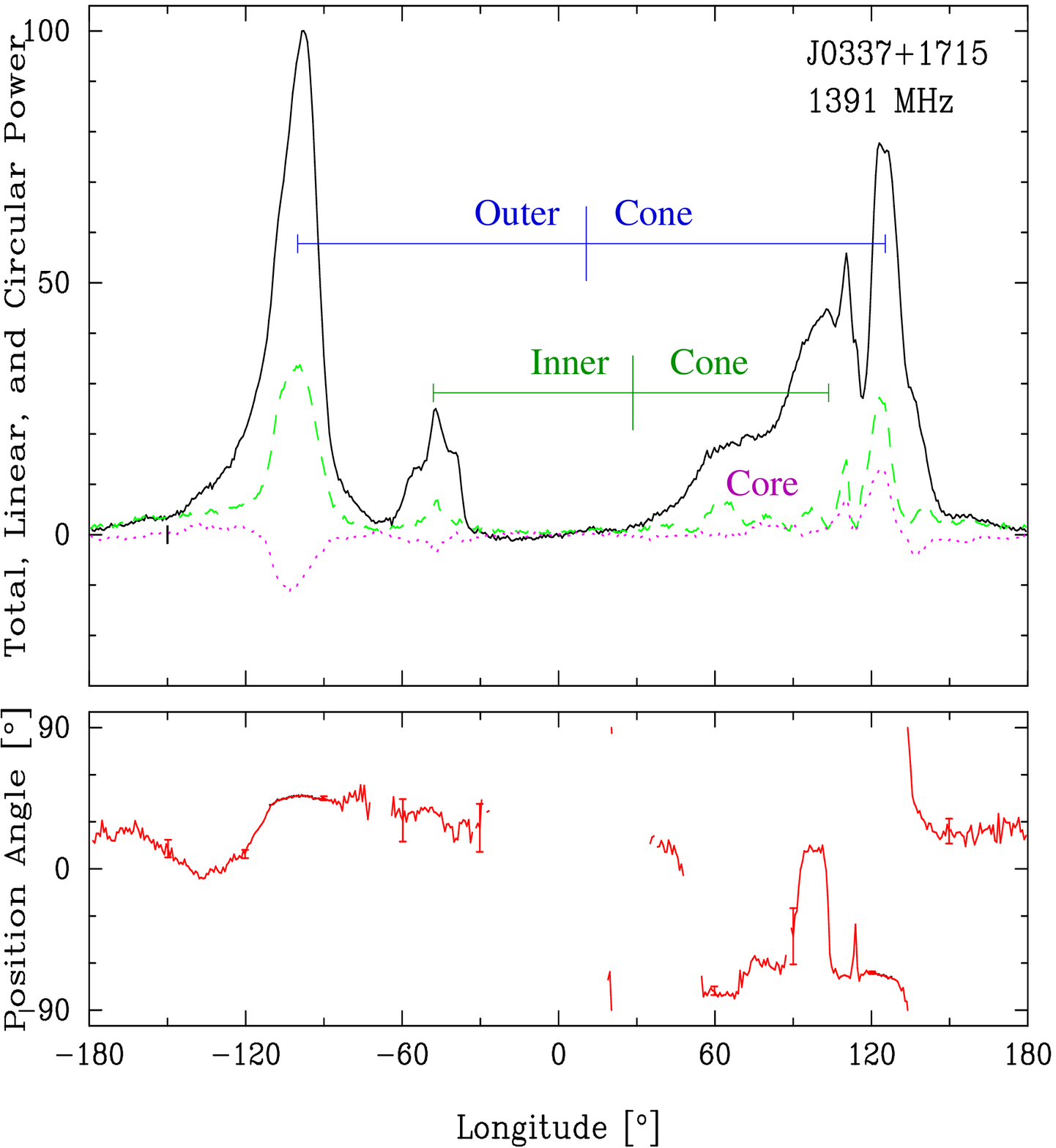}}
\caption{Total 1.4-GHz Arecibo PUPPI polarized profile of pulsar J0337+1715 showing 
its apparent core component as well as its pairs of inner and outer conal components.  
The core component seems to be conflated with the trailing conal components at about 
90\degr\ longitude.  The positions and centers of the two conal component pairs are 
indicated.  They are far from symmetrically spaced as in slower pulsars, apparently 
because of aberration/retardation.  However, note that the outer cone center precedes 
the inner one as well as the core.  The upper panel gives the total intensity (Stokes $I$; 
solid curve), the total linear ($L$ [=$\sqrt{Q^2+U^2}$]; dashed green), and the circular 
polarization (Stokes $V$; dotted magenta).  The PPA [=$(1/2)\tan^{-1} (U/Q)$]  values 
(lower panel) have been Faraday-derotated to infinite frequency.  Errors in the PPAs 
were computed relative to the off-pulse noise phasor---that is, $\sigma_{PPA} \sim tan^{-1 } 
(\sigma_\mathrm {off-pulse}/L)$ and are plotted when $<30\degr$ and indicated by 
occasional 3-$\sigma$ error bars.}
\label{fig1}
\end{center}
\end{figure}

The profile of millisecond pulsar J0337+1715 surprisingly has what appears to be a 
five-component core/double-cone configuration.  Such a five-component structure is 
potentially important because it has not been clear whether millisecond pulsars ever 
exhibit emission-beam configurations similar to those of slower pulsars.  J0337+1715 
is now well known as the unique example so far of a millisecond pulsar in a system of 
three stars (Ransom \etal\ 2014).  The pulsar is relatively bright and can be observed 
over a large frequency band, so that both its pulse timing and polarized profile 
characteristics can be investigated in detail.  

In addition, we find three other probable examples of MSPs with core-cone profiles:  
the original binary pulsar, B1913+16; the second MSP discovered, B1953+29; and 
pulsar J1022+1001.  We will also discuss the profile geometries of these three MSPs 
and consider their interpretations below.  

The slow pulsars with four-component profiles show outer and inner pairs of conal 
components, centered closely on the (unseen) longitude of the magnetic axis, all 
with particular dimensions relative to the angular size of the pulsar's magnetic 
(dipolar) polar cap.  The classes of pulsar profiles and their quantitative beaming 
geometries are discussed in Rankin (1983, ET I) and ET VI, and other population 
analyses have come to similar conclusions (\eg Gil \etal\ 1993; Mitra \& Deshpande 
1999; Mitra \& Rankin 2011).  Core beams have intrinsic half-power angular diameters 
about equal to a pulsar's (dipolar) polar cap ($2.45\degr P^{-1/2}$, where $P$ is 
the pulsar rotation period), suggesting generation at altitudes just above the 
polar cap.  Conal beam radii in slow pulsars also scale as $P^{-1/2}$ and reflect 
emission from heights of several hundred km.  Core emission heights are not 
usually known apart from the low altitude implication of their widths---so it is 
plausible to assume that core radiation arises from heights well below that of the 
conal emission.  

Classification of profiles and quantitative geometrical modeling provides an essential 
foundation for understanding pulsars physically.  MSPs, heretofore, have seemed 
inexplicable in these terms.  The core/double-cone model works well for most pulsars 
with periods longer than about 100 ms, whereas faster pulsars mostly either present 
inscrutable single profiles or complex ones where no cones or core can be identified.  
It is then important to learn how and why it is that the core/double-cone beam model 
often breaks down for the faster pulsars.  

Aberration/retardation (hereafter A/R) becomes more important for faster pulsars 
and MSPs simply because their periods are shorter compared to A/R temporal shifts.  
The role of A/R in pulsar emission profiles was first described by Blaskiewicz \etal\ 
(1991, hereafter BCW), but only later was this understanding developed into a 
reliable technique for determining pulsar emission heights (\eg Dyks \etal\ 2004; 
Dyks 2008; Mitra \& Rankin 2011; Kumar \& Gangadhara 2012).  A/R has the 
effect of shifting the profile earlier and the polarization-angle (hereafter PPA) 
traverse later by equal amounts 
\begin{equation}
	\Delta t =\pm2h_{em}/c
		\label{eq1}				
\end{equation}
where $h_{em}$ is the emission height (above the center of the star), such 
that the magnetic axis longitude remains centered in between.  In 
that the PPA inflection is often difficult to determine, A/R heights have also been 
estimated by using the core position as a marker for the magnetic axis longitude 
as first suggested by Malov \& Suleimanova (1998) and Gangadhara \& Gupta 
(2001) and now widely used.  The A/R method has thus usually been used to 
measure conal emission heights, but core heights have also been estimated 
for B0329+54 (Mitra \etal\ 2007) and recently for B1933+16 and several other 
pulsars (Mitra \etal\ 2016) giving heights around or a little lower than the conal 
ones.   Crucially, A/R measurements indicate emission heights of some 300-500 
km in the slow pulsar population.  So it is important to see how MSPs compare, 
given that many have magnetospheres smaller than 500 km.  

In what follows we discuss the observations in \S\ref{obs}, the J0337+1715 core 
component and its width in \S\ref{ccw}, and the conal component configuration in 
\S\ref{dcc}.  \S\ref{qg} treats its quantitative geometry,  \S\ref{a/r} computes 
aberration/retardation emission heights, and \S\ref{msps} considers three other 
millisecond pulsars with cone/single-cone profiles.  Finally, \S\ref{sum} summarizes 
the results,  \S\ref{phys} interprets them physically, and \S\ref{con} outlines the 
overall conclusions.

\begin{table}
\begin{center}
\caption{Properties of the Arecibo and GBT observations.}
\begin{tabular}{lcccc}
\hline
\hline
Band &Backend & MJD &Resolution & Length\\
(GHz) &    && (\degr/sample)  & (s)\\
\hline
\multicolumn{5}{l}{{\bf J0337+1715}  ($P$=2.73 msec; $DM$=21.32 pc/cm$^3$)} \\
 1.1-1.7 &PUPPI& 56736 & 0.70 & $\sim$3400 \\
 1.1-1.8 &GUPPI& 56412 & 0.70 & 40968 \\
 0.42-0.44 &PUPPI& 56584 & 1.41 & $\sim$3400 \\
 1.9-2.5* &PUPPI& 56781 & 1.41 & 3600 \\
 1.9-2.5 &PUPPI& 57902 & 1.41 & 3600 \\
 2.5-3.1* &PUPPI& 56768 & 1.41 & 3936 \\
 2.5-3.1 &PUPPI& 57902 & 1.41 & 3000 \\
 1.1-1.7 &Mocks& 56760 & 4.74 & 1000 \\
 
\multicolumn{5}{l}{{\bf B1913+16}  ($P$=59.0 msec; $DM$=168.77 pc/cm$^3$)} \\
  1.1-1.9 & GUPPI& 55753 & 0.35 & 1900 \\
  1.1-1.7 &Mocks& 56199 & 0.58 & 1200 \\

\multicolumn{5}{l}{{\bf B1953+29}  ($P$=6.13 msec; $DM$=104.501 pc/cm$^3$)} \\
  1.1-1.6 &Mocks& 56564 & 2.24 & 2400 \\
  1.1-1.6 &Mocks& 56585 & 2.24 & 2400 \\
  1.1-1.9 &PUPPI& 57129 & 0.35 & 3608 \\
  
\multicolumn{4}{l}{{\bf J1022+1001}  ($P$=16.4 msec; $DM$=10.2521 pc/cm$^3$)}\\
  1.1-1.7 &Mocks& 56315/23 & 0.53 & 600 \\
  1.1-1.7 &Mocks& 56577 & 0.46 & 1800 \\
  0.30-0.35 &Mocks& 56418/31 & 2.55 & 600 \\
  0.30-0.35 &Mocks& 56577 & 2.55 & 1800 \\
\hline
\end{tabular}
\end{center}
\label{tab1}
\footnotesize
Notes:  These earlier PUPPI 2.0- and 2.8-GHZ observations could not be 
calibrated polarimetrically because of faulty cal signals; therefore, we use only 
Stokes $I$ in Figure~\ref{figA1}.
\end{table}

\section{Observations}
\label{obs}
\subsection{Acquisition and calibration}
\noindent {\bf J0337+1715} A large number of observations at different frequencies 
and observatories were carried out in the course of confirming J0337+1715 and 
understanding the complex orbital modulations of its pulse-arrival times.  Some 
of these were polarimetric and some much more sensitive, and unsurprisingly the 
Arecibo observations were usually the highest quality.  In compiling the observations 
for this analysis, we assessed which observations were deepest at both 1400 and 
430 MHz.  The PUPPI backend{\footnote{http://www.naic.edu/{$\sim$}astro/guide/node11.html}}
was used at both frequencies with a 30-MHz bandwidth at 430 and a 600-MHz 
bandwidth at 1400 MHz.  We then explored the pulsar's behavior at higher frequencies, 
and both 2.0 and 2.8-GHz observations are reported here.  We also attempted 
a rise-to-set observation at 5 GHz with a 1-GHz bandwidth, but the pulsar was not 
detected.  All the J0337+1715 observations were calibrated or re-calibrated using 
the PSRchive {\rm pac -x} routine (Hotan, van Straten \& Manchester 2004).  
The final bin size was chosen to reflect the joint time and frequency resolution, and 
is given in Table~\ref{tab1} along with some other characteristics of the observations.   

\begin{figure}
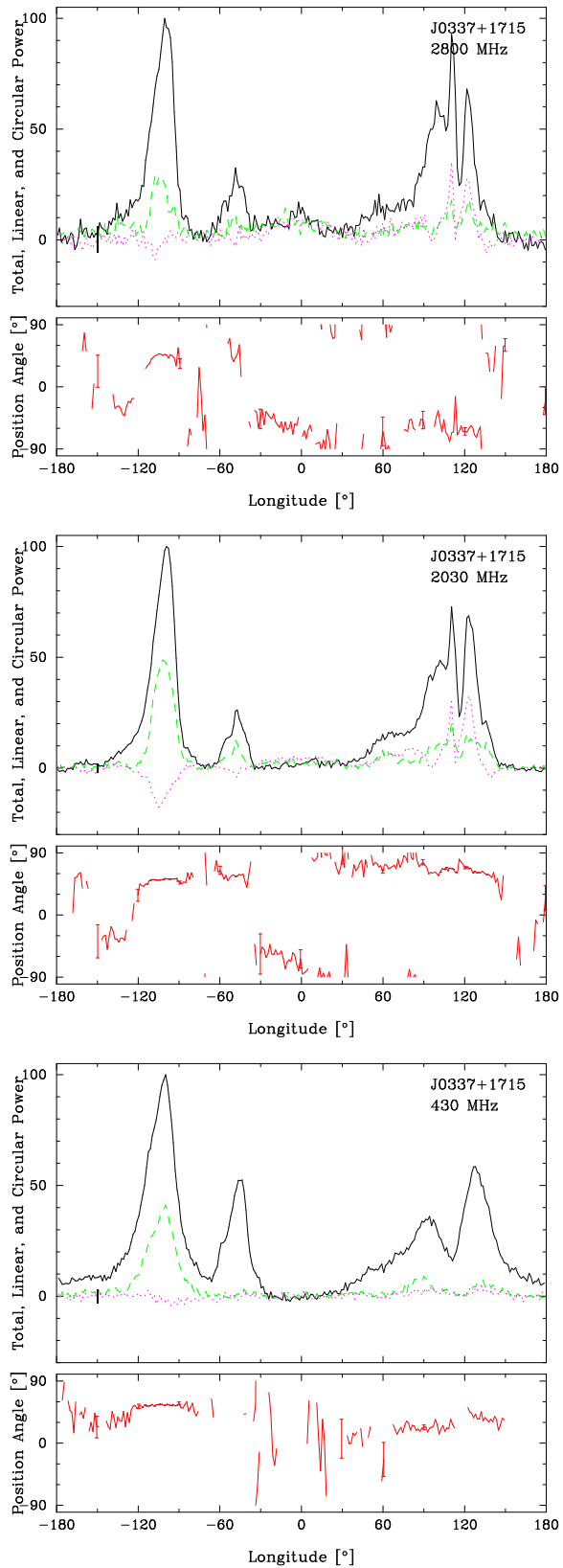

\begin{center}
\mbox{\includegraphics[width=70mm,angle=-90.]{PQpJ0337+1715.57900su_rot.ps}} 
\vskip 0.3cm
\mbox{\includegraphics[width=70mm,angle=-90.]{PQpJ0337+1715.57900sl_rot.ps}} 
\vskip 0.3cm
\mbox{\includegraphics[width=70mm,angle=-90.]{PQpJ0337+1715u_shifted_PArot.ps}}
\caption{Arecibo PUPPI profiles of J0337+1715 after Fig.~\ref{fig1} at higher and lower 
frequencies than the earlier 1391-MHz profile; see Table~\ref{tab1}.  These show the 
near absence of evolution with frequency.  The profiles, as well as that at 1.4-GHz above, 
are aligned with the dispersive delay removed.}
\label{fig3}
\end{center}
\end{figure}

The 1391-MHz profile in Fig.~\ref{fig1} is the deepest and best resolved of the four 
and deserves more detailed analysis.  The 2.8, 2.0 and 0.43 GHz profiles can be seen 
in Fig.~\ref{fig3}. 

\noindent {\bf B1913+16, B1953+29 and J1022+1001.}  We also carried out Arecibo 
observations of these pulsars using the L-band Wide feed and the Mock{\footnote 
{http://www.naic.edu/{$\sim$}astro/mock.shtml}} spectrometers.  The former were single-pulse 
polarimetric observations using as many Mocks as needed to optimize the resolution 
and use the total available bandwidth for maximal sensitivity, and the results are given
in Table~\ref{tab1}.  These Mock observations were processed as described in Mitra 
\etal\ (2016) including derotation to infinite frequency.

\subsection{Rotation-measure determinations}
Our rotation-measure determinations will be published as a part of a larger paper 
in preparation together with a full description of our observations and techniques.  

\noindent {\bf J0337+1715.}  A rotation measure ($RM$) of +30$\pm$3 rad-m$^2$ 
was determined using a set of the best quality observations from both the Arecibo 
PUPPI and Green Bank Telescope (hereafter GBT) GUPPI machines\footnote{https://safe.nrao.edu/wiki/bin/view/CICADA/GUPPiUsersGuide}.  These were processed as above with 
PSRchive, and its {\rm rmfit} routine was used to estimate each $RM$ and its error; 
the stated error then reflects the scatter of these values.  Ionospheric corrections 
estimated for these observations ran between +0.8 and --0.2 rad-m$^2$, so the 
intrinsic value lies well within the above error.  Separately, the Mock observation of 
MJD 56760 was used to determine an ionosphere-corrected value of +29.3$\pm$0.7 
rad-m$^2$.

\noindent {\bf B1913+16.}  An accurate $RM$ value was determined for the 
first time using five Arecibo Mock 1.4-GHz observations---one of which is shown 
below in Fig.~\ref{fig5}.  After ionospheric correction the $RM$s were estimated by 
trial and error maximization of the aggregate linear polarization resulting in a value 
of +354.4$\pm$0.6 rad-m$^2$, where the error reflects the {\it rms} scatter of the 
values.  This value is then further confirmed by a Green Bank Telescope (hereafter 
GBT) GUPPI observation, made as part of another project (Force \etal\ 2015) and 
processed using the PSRchive {\rm rmfit} routine to yield an $RM$ value of 
+357.9$\pm$1.5 rad-m$^2$ that included some 2-3 units of ionospheric contribution.  
This represents a substantial increase in precision compared to the current value 
on the ATNF Pulsar Catalog website of +430$\pm$77 rad-m$^2$ (Han \etal\ 2006).

\noindent {\bf B1953+29.}  This $RM$ value was also estimated for the first time 
using Arecibo 1.4-GHz observations as processed by both the Mock spectrometers 
and PUPPI.  The two Mock observations were processed as above and together 
yielded a value of +3.0$\pm$0.4 rad-m$^2$, corrected for ionospheric contributions;
the second of the two is shown in Fig.~\ref{fig6}.  In addition, a PUPPI observation was 
processed as above using PSRchive {\rm rmfit} and yielded a value of +5.7$\pm$3.0 
rad-m$^2$ which was not corrected for the expected 2-3 units of ionospheric RM.

\noindent {\bf J1022+1001.}  This pulsar has a well determined $RM$ value of 
+1.39$\pm$0.05 rad-m$^2$ on the ATNF Pulsar Catalog website due to Noutsos \etal\ 
(2015).

\section{Exploration of the J0337+1715 profile}
\subsection{The Putative Core Feature and its Width}
\label{ccw}
In slower pulsars, the core component width reflects the angular diameter of the 
polar cap near the stellar surface, and as such has an intrinsic half-power diameter 
of $2.45\degr P^{-1/2}$ ($P$ is the stellar rotation period), but the observed width 
entails a further factor of $\csc\alpha$ (where $\alpha$ is the colatitude of the 
magnetic axis with respect to the rotation axis).  The expected intrinsic width for 
this 2.73-ms MSP would then be some 47\degr\ and the observed width 67.5\degr, 
given that $\alpha$ is plausibly 44\degr\ (Ransom \etal\ 2014) if the orbital and 
rotational angular momenta are aligned.  In J0337+1715's profile, the putative 
core is conflated with the trailing inner conal component over the entire observable 
band, therefore the core width can only then be estimated.  Exploratory modeling 
of the four profiles in the Appendix suggests a core width of between 60 and 
70\degr, so we have fixed the core width in our modeling at the above expected 
67.5\degr\ observed value.

\subsection{Double Cone Configuration}
\label{dcc}
In slower pulsars, conal components occur in pairs for interior sightline traverses 
where the sightline impact angle $|\beta|$ is smaller than the conal beam radius 
$\rho$.  Geometrically, these conal component pairs are found to be of two types, 
inner and outer cones, that have specific (outside, half-power, 1-GHz) radii $\rho$ 
of $4.33\degr P^{-1/2}$ and $5.75\degr P^{-1/2}$ (ET VIa,b; see VIa eq.(4)).  
In pulsars where both cones and a core are observed, profiles then have five 
components, such as in pulsar B1237+25.  

We suggest that pulsar J0337+1715's profile in Fig.~\ref{fig1} may exhibit just these 
five components as well as a narrow (putative caustic; see below) feature on 
the outer edge of the trailing inner conal component, and we also model these 
features in the Appendix.  The leading outside (LOC) and inside conal (LIC) 
components are seen to fall at about --100\degr\ and --50\degr\ longitude, 
respectively.  

The core is conflated with the trailing conal components (TIC and TOC) at some 
+100\degr\ and +125\degr, as modeled and better estimated in Table~A1.  
The conal components are asymmetric and the rightmost one has a trailing edge 
bump.  However, the Appendix modeling appears to locate their centers and half 
widths adequately.  The ``spike'' on the trailing edge of the trailing inner conal 
feature above 1 GHz deserves special mention, as it seems to resemble the 
``caustic''\footnote{``Caustic'' refers to a field-line geometry in which an 
accidentally favorable curvature tracks the sightline producing a bright narrow 
broadband feature (\eg Dyks \& Rudak 2003).} features that are seen in some 
other pulsars.  The 1391-MHz polarization-angle (PPA) traverse shows what 
seems to be a 90\degr\ ``jump'' at about 135\degr\ longitude, and earlier ones 
at about 50\degr\ and 100\degr; if these are resolved, then the PPA seems to rotate 
rather little under the components as seems possibly the case at the other 
more depolarized frequencies.

\subsection{Quantitative Geometry}
\label{qg}
Profiles for frequencies both higher and lower than that in Fig.~\ref{fig1} are 
given in Figure~\ref{fig3}.  Note that the pulsar's five components can easily be 
discerned in these profiles up to 2.8 GHz and down to 430 MHz.  The conal 
features of the 430-MHz profile are broader and appear less well resolved 
compared to the higher frequency profiles; however, this appears not to be 
the result of poorer instrumental resolution and scattering.  The core is 
obviously conflated with the trailing inner conal component across all the 
observations, and this must be taken into account in identifying the structures 
contributing to the pulsar's overall profile.  

The centers (C), widths (W) and amplitudes (A) of J0337+1715's components 
are given in Table~A1, where an effort was made to determine these values 
by fitting methods.  However, given the irregular shapes of the components, 
the fitting errors are small compared to systematic ones of a few degrees.  The 
table thus gives the positions and widths of the conal components so obtained 
as well as those of the putative ``caustic'' feature (CC in Table~A1) 
whose center could be fitted within about 1\degr---and it is satisfying to see 
that this feature aligns in the high frequency profiles within this latter error.  
Unsurprisingly, the core component properties are more difficult to estimate 
in the various profiles, even when taking the widths as fixed at the expected 
value as shown in the table.  We discuss this further below in connection with 
the A/R analysis where this difficulty is most pertinent.  

The properties of the two cones are given in Table~\ref{tab2}, where $w_i$ 
and $w_o$ are the inner and outer outside, half-power conal widths, computed 
as the longitude interval between the component-pair centers plus half the 
sum of their widths.  Here we see that the inner and outer conal widths are 
essentially constant over the observable frequency band.  This near constancy 
of profile dimensions over some three octaves is similar to that observed for 
other millisecond pulsars where there is little change in the components 
separations over the total band of observations (\eg Kondratiev \etal\ 2016).  

Here we apply the standard spherical geometric analyses assuming a central 
sightline traverse (sightline impact angle $\beta$ about 0\degr\ as suggested by 
the relatively constant PPA traverse), that the magnetic latitude $\alpha$ is 44\degr, 
and that the emission occurs adjacent to the ``last open field lines'' (\eg ET VIa).  
This analysis is summarized in Table~\ref{tab2}, where $\rho_i$ and $\rho_o$ 
are the computed conal emission beam radii (to their outside half-power 
points) per eq.(4) in ET VIa.  The emission characteristic heights are then 
computed assuming bipolarity using eq.(6).  

The outside half-power conal radii of about 54\degr\ and 75\degr\ are of course 
enormous compared to those found for slow pulsars.  However, they are substantially 
smaller---only about 2/3---what would be expected according to the slow pulsar 
relationships $4.33\degr P^{-1/2}$ and $5.75\degr P^{-1/2}$ for the outside 3-dB 
radii of inner and outer cones.  The model emission heights of some 53 and 100 km 
are also correspondingly smaller---less than half of the 120 and 210 km typically 
seen in the slow pulsar population.  However, it is important to recall that these 
are {\em characteristic} emission heights, not physical ones, estimated using the 
convenient but problematic assumption that the emission occurs adjacent to the 
``last open'' field lines.  It will be interesting to compare these heights with the 
physical emission heights estimated using aberration/retardation (A/R) just below.  

\begin{table}
\begin{center}
\caption{Double Cone Geometry Model for PSR J0337+1715}
\begin{tabular}{c c c c c c c}
\hline\hline
Freq   &   $w_i$  &$\rho_i$  &$w_o$   &$\rho_o$  & $h_i$ & $h_o$  \\
(MHz) &(\degr)&(\degr) &(\degr) &(\degr)  &(km)   & (km)   \\
\hline
2800 & 162 & 54 & 238 & 74 & 52 & 99  \\
2080 & 163 & 54 & 242 & 74 & 53 & 101  \\
1391 & 166 & 55 & 240 & 74 & 53 & 99  \\
 430  & 163 & 54 & 252 & 76 & 53 & 107  \\
\hline
\hline
\label{tab2}
\end{tabular}
\end{center} 
\footnotesize
Notes:  $w_i$ and $w_o$ are the outside half-power inner and outer conal 
component widths scaled from Figs.~\ref{fig1} and \ref{fig3}.  $\rho_i$ and $\rho_o$ 
are the outside half-power conal beam radii computed from ET Paper VIa, eq.(4), 
and $h_i$ and $h_o$ are the respective geometric conal emission heights 
computed from the latter paper's eq.(6) assuming bipolarity and emission along 
the polar fluxtube boundary.
\end{table}

\begin{table*}
\begin{center}
\caption{Aberration/Retardation Analysis Results for PSR J0337+1715}
\begin{tabular}{rccccccc}
\hline
  Freq/Cone  & $\phi^i_l$ & $\phi^i_t$ & $\phi^i_c $ & $\kappa^i$ & $\rho^i$ & $h^i_{em}$ & $s^i$ \\
  (MHz)   &  (deg) & (deg) & (deg)  & (deg)& (deg) & (km) &  \\
\hline
\hline
\\
2800/II & $-105\pm3$&$124\pm4$&$10\pm5$&$-75\pm11$&$71.5\pm1.6$&$86\pm13$&$0.91\pm0.07$ \\
2030/II & $-102\pm3$&$125\pm4$&$11\pm5$&$-74\pm11$&$71.1\pm1.7$&$84\pm13$&$0.92\pm0.07$ \\
1391/II & $-100\pm3$&$128\pm4$&$14\pm5$&$-71\pm11$&$71.2\pm1.6$&$81\pm13$&$0.94\pm0.07$ \\
430/II   & $-103\pm3$&$129\pm4$&$13\pm5$&$-72\pm11$&$72.2\pm1.6$&$82\pm13$&$0.94\pm0.07$ \\
\\
2800/I & $-51\pm2$&$98\pm4$&$24\pm4$&$-62\pm11$&$49.5\pm1.9$&$70\pm12$&$0.74\pm0.07$ \\
2030/I & $-49\pm2$&$97\pm4$&$24\pm4$&$-61\pm11$&$48.9\pm1.9$&$69\pm12$&$0.74\pm0.07$ \\
1391/I & $-47\pm2$&$99\pm4$&$26\pm4$&$-59\pm11$&$48.9\pm1.9$&$67\pm12$&$0.75\pm0.07$ \\
430/I   & $-48\pm2$&$94\pm4$&$23\pm4$&$-62\pm11$&$47.5\pm2.0$&$71\pm13$&$0.71\pm0.06$ \\
\hline
\label{tab3}
\end{tabular}
\end{center} 
\footnotesize
Notes: $\phi^i_l$ and $\phi^i_t$ are the leading and trailing conal component positions; 
$\phi^i_c$ their centers; $\kappa^i$ are the total A/R shifts in longitude as marked by the core 
component centers; $h^i_{em}$ the A/R emission heights computed from eq.(\ref{eq2}); and 
$\rho^i$ and $s^i$ are the conal radii and emission annuli within the polar fluxtube computed 
using ET Paper VIa eq.(4) and $\sin(2\rho^i/3)\sqrt{R_{lc}/h^i}$, respectively.  
\end{table*}

\subsection{Aberration/Retardation Analysis}
\label{a/r}
Here we see that the centers of the outer and inner conal component pairs precede 
the longitude of the core-component center by some 60\degr\ or more---and we propose 
to interpret this shift as due to aberration/retardation (A/R) to assess whether the 
results are appropriate and reasonable.  A difficulty is determining the position of 
the core component accurately, given its conflation with the outer conal components.   
Table~A1 below reports the results of modeling the pulsar's features 
with Gaussian functions in order to assess the character of this conflation and then 
estimate the core component's position and center.   Assuming a constant core width 
at its expected 67.5\degr\ value and a constant amplitude in relation to the four conal 
components, the core center falls close to --85\degr\ in each band within about 2\degr.  
However, other modeling assumptions might well yield somewhat different values.  
We therefore propose to take the core center at --85\degr$\pm$10\degr\ in order to 
incorporate the major range of estimates and uncertainties.  Finally, as emphasized 
above, core components exhibit a predictable width in slow pulsars that reflects both 
the polar cap geometry and the magnetic colatitude.  This is the rationale for holding 
the core width here to its expected value.  If, however, J0337+1715, for whatever 
reason, had a core that was smaller or larger, then the center would shift by half the 
difference amount.

Table~\ref{tab3} below gives A/R analyses for the two cones in the four bands.  
The column values give the peaks of the leading and trailing conal components 
$\phi_l$ and $\phi_t$, their center point $\phi_c$=($\phi_t+\phi_l$)/2, and the offsets.  
The total A/R shift is then given by $\kappa$, which is computed as the longitude 
interval between the conal centers $\phi_c$ and the longitude of the magnetic axis 
as marked by the core component, or  +85\degr\ -- $\phi_c$, and its uncertainty is 
dominated by the above $\pm$10\degr\ in the estimated core position.  

The conal radii $\rho$ corresponding to the conal component-pair peak separations 
are calculated geometrically as in Table~\ref{tab2}, and those values are somewhat 
larger as expected, about 54 and 75\degr, than tabulated here, given their reference 
to the outside half-power points rather than the component peaks.  This computation 
is independent of the A/R height determination and is used only to estimate the annuli 
of the emitting locations $s^i$, where 0 lies along the magnetic axis and unity 
on the polar flux tube boundary.  That these values for the outer cone are close to 
unity tends to support our geometrical model assumption above that its emission 
regions lie on the periphery of the (dipolar) polar fluxtube; the inner cone is then 
emitted more to the interior and higher in altitude.  

In A/R analysis here per eq.({\ref{eq2}}) the relation 
\begin{equation}
	h_{em} = -c\phi_c P/360\degr/2
	\label{eq2}
\end{equation}
provides a 
reliable estimate of the emitting region height, subject only to correct interpretation 
of the core-cone beam structure and determination of the core component's position 
as marking the magnetic axis longitude or a region just above it---in any case at a 
much lower altitude than the conal emission.  The resulting physical A/R emission 
heights for the inner and outer cones are then about 70 km and 85 km, respectively, 
where the velocity-of-light cylinder distance is $R_{lc}$ (=$cP/2\pi$) of 130 km.  
We find no significant emission-height differences over the nearly three octaves of 
observations.  This analysis parallels those for slower pulsars in Force \& Rankin 
(2010), Mitra \& Rankin (2011), Smith \etal\ (2013) and most recently Mitra \etal\ 
(2016) and are based on BCW as corrected by Dyks \etal\ (2004).

\section{Apparent Core-Cone Beam Structure in Other MSPs}
\label{msps}

\subsection{B1913+16}

\begin{figure}
\begin{center}
\mbox{\includegraphics[width=63mm,angle=0.]{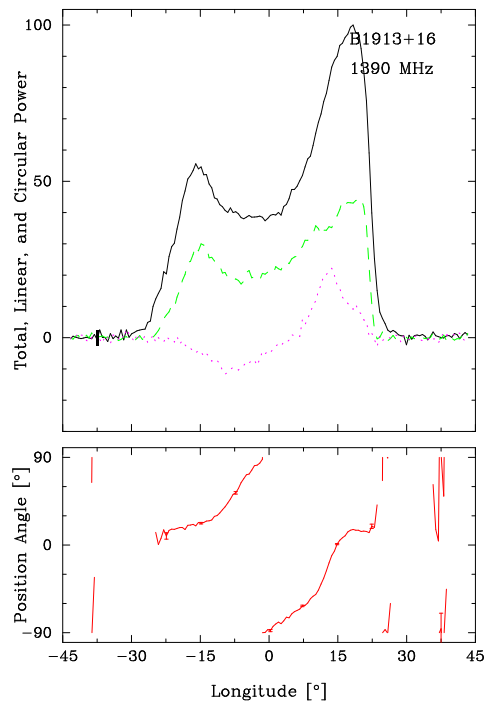}}
\caption{Arecibo 1.4-GHz Mock-spectrometer profile of B1913+16 from 2012 
September 29.  The form of this profile differs substantially from those measured 
earlier (\eg BCW) due to the star's precession.  This observation was carried out 
with four Mocks  sampling 86-MHz bandwidths and 95 $\mu$sec resolution.  The 
PPAs have been derotated to infinite frequency using the measured $RM$ value 
above, ionosphere corrected, of +354.4$\pm$0.4 rad-m$^2$.}
\label{fig5}
\end{center}
\end{figure}

\begin{table}
\begin{center}
\caption{Conal Geometry Models for B1913+16, B1953+29 and J1022+1001}
\begin{tabular}{l c c c c c c c}
\hline\hline
Pulsar &$P$& $\alpha$ &$\beta$ & $w$  &$\rho$ & $h$   \\
            & (ms) &(\degr)&(\degr) &(\degr) &(\degr) &(km)  \\
\hline
B1913+16 &59.0& 46 & 4.0 & 47 & 17 & 126  \\
B1953+29 &6.1& $>$65 & --18  & 100   & 44.4  & 81  \\
J1022+1001 &16.4& $\sim$60 & $\sim$7& 42 & 20 & 45  \\
\hline
\hline
\label{tab4}
\end{tabular}
\end{center} 
Notes:  $P$, $\alpha$ and $\beta$ are the pulsar's rotation period, magnetic 
colatitude and sightline impact angle, respectively.  The $w$s are the outside 
half-power conal component widths measured from the fitting of Figs.~\ref{figA3}
\ref{figA4} and \ref{figA5}.  The $\rho$s are the outside half-power conal beam 
radii computed from ET Paper VIa, eq.(4), and the $h$s are the respective 
geometric conal emission heights computed from the above eq.(6) assuming 
bipolarity and emission along the polar fluxtube boundary.
\end{table}

The Binary Pulsar B1913+16 has been studied intensively using timing methods, 
resulting in the identification of gravitational radiation (\eg Weisberg \& Huang 
2016) but until recently the pulsar was difficult to study polarimetrically.  Moreover, 
the star precesses causing secular changes in its profile and polarization 
(\eg Weisberg \& Taylor 2002).  However, its early basic profile morphology had 
been known to  be tripartite at 430 MHz and double at 1400 MHz and above.  
This is just as seen in many slow pulsars, and one of us classified this pulsar 
as having a core/inner-cone triple ({\bf T}) profile (ET Paper VIb) on the basis of 
the 430- and 1400-MHz profiles then recently published by Blaskiewicz \etal\ 
(1991; their fig. 17).  These profiles show a typical evolution due to the core's 
relatively steep spectrum---such that in B1913+16 the core is not resolvable at 
frequencies above 1 GHz.  Key to understanding the central component as 
being a core is its width, which reflects the angular size of the star's polar cap 
at $2.45\degr P^{-1/2}$ or 10\degr\ intrinsically and then broader by a factor 
of $\csc\alpha$ in profile longitude.  Our Gaussian fit to the 1990 430-MHz 
profile in Fig.~\ref{figA3} gives a value of 17$\pm$2\degr\ which supports both the 
core identification and an $\alpha$ value of about 45\degr\ as in ET VI (where 
the 1-GHz width is taken as 14\degr.  This in turn is quite reasonable if the star's 
spin and orbital rotations were aligned, but the pulsar's precession shows that 
they are not.  The star's changing profile forms at 430 MHz suggest a core/cone 
beam moving toward and ever less central sightline traverse, greatly complicating 
models of its emission geometry.\footnote{Some precession models suggest 
smaller $\alpha$ values (Kramer 1998; Weisberg \& Taylor 2002; Clifton \& Weisberg 
2008), but both the core and conal widths seem too narrow to support this.  Such 
models incur their own assumptions and difficulties such that the differences 
between methods are not easily reconciled.} 

We then take the geometric model in Paper ET VIb, table 4, based on BCWs 
profiles---at a time when the 430-MHz core was most clearly seen---as close to 
the mark for our purposes here.  Its central sightline traverse makes it insensitive 
to $\beta$.  Similarly, the 430-MHz profile is broader at the conal outside 
half-power points, but the conal peak separation is nearly constant at about 
40\degr\ between the two frequencies, so the increased breadth must be intrinsic 
and the cone thus an inner one.  Therefore, we confirm the core/inner-cone 
geometry of Paper VI.  

BCW pioneered the use of A/R for determining pulsar emission heights, and their 
analysis for B1913+16 is illustrated in their fig. 17.  They determined the A/R shift 
from the PPA steepest gradient point (SG) backward to the center of the conal 
component pair.  Arrows in the figures show these points, and according to their 
analysis the A/R shift, averaged between the two frequencies, is 12.5\degr$\pm$
1.0\degr, that then corresponds to an emission height of 126$\pm$21 km.  This 
in turn then occurs within a light-cylinder radius of 2820 km.  

If the above two points are well identified, the above A/R value follows the method 
BCW advocated.  We can then try to estimate the emission height by two other 
methods:  first, the fits in Fig.~\ref{figA3} clearly show that the core component 
peak lags the center point between the two conal peaks, and this lag is 1.5\degr\ 
or 0.246 msec or some 37 km in altitude per eq.(\ref{eq1})---that is, the conal 
emission region appears to be higher than the core region by this amount.  Second, 
the zero-crossing point of antisymmetric circular polarization often marks the 
longitude of the magnetic axis,{\footnote{Core components in slow pulsars often 
exhibit antisymmetric circular polarization, such that the zero-crossing point marks 
the longitude of the magnetic axis.  No core is resolvable in Fig.~\ref{fig5}, but the 
``bridge'' region between the two conal peaks is probably dominated by core 
emission.} and in the B1913+16 profile of Fig.~\ref{fig5} this point falls 6$\pm$1\degr\ 
or some 1.0 msec after the conal center point suggesting a conal emission height 
of 148$\pm$25 km.  As the magnetic axis longitude falls halfway between the 
oppositely shifted profile and PPA traverse under A/R, the latter result is compatible 
with BCW's interpretation and the 37-km shift may represent the height between the 
core and conal emission regions.

\subsection{B1953+29}
\begin{figure}
\begin{center}
\mbox{\includegraphics[width=70mm,angle=-90.]{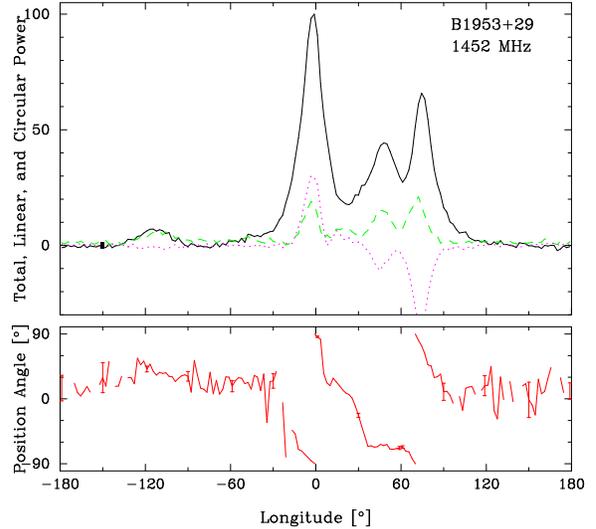}}
\caption{Arecibo 1.4-GHz Mock-spectrometer profile of B1953+29 from 2013 October 20.  
This observation was carried out with seven Mocks sampling adjacent 34-MHz bands 
with a 38 $\mu$sec sampling time.  Note what appears to be an interpulse preceding 
the putative core-cone triple ({\bf T}) main pulse.   We believe this is the first observation 
to clearly show the pulsar's PPA traverse and interpulse.  The PPAs have been derotated 
to infinite frequency using its measured RM of +2.9$\pm$0.6 rad-m$^2$ (including some 
0.8 rad-m$^2$ in the ionosphere).  A further Arecibo observation carried out with PUPPI 
confirms the profile structure and $RM$ value.}
\label{fig6}
\end{center}
\end{figure}

The second millisecond pulsar, B1953+29, aka Boriakoff's Pulsar (Boriakoff \etal\ 
1983) is also a binary pulsar with a 117-day orbit.  It has also had little subsequent 
study as few profiles have been measured in the intervening decades.  Profiles 
and provisional polarimetry at both 430 and 1400 MHz were presented at an NRAO 
workshop just after the discovery (Boriakoff \etal\ 1984) but not otherwise published, 
and the polarimetry efforts appearing since (Thorsett \& Stinebring 1990; Xilouris 
\etal\ 1998; Han \etal\ 2009; Gonzalez \etal\ 2011) leave many questions unanswered.  
The pulsar was difficult to observe polarimetrically, but Arecibo's L-band Wide feed 
together with the Mock spectrometers or PUPPI now permit much more sensitive and 
resolved observations.  

Boriakoff \etal's 430-MHz profile is single, showing only weak inflections from 
power in conal outriders on both sides of the broad and relatively intense central 
component.  At the higher frequency, however, all three components are visible and 
distinct.  This is just the evolution seen for most conal single ({\bf S}$_t$) pulsars as 
first described in Rankin (ETI, ETVIa).  No full beamform computation was given for 
the pulsar in ETVIb because no reliable polarimetry was then available.  

Figure~\ref{fig6} shows a recent Arecibo 1452-MHz polarimetric observation 
conducted with the Mock spectrometers.  It is far more sensitive and better resolved 
than any previously published polarimetry for this important pulsar, and clearly 
shows its PPA traverse as well as what appears to be a bright interpulse.\footnote{This 
interpulse is also clearly seen in the timing study of Gonzalez \etal\ (2011)---and 
hinted at in some of the earlier profiles---and their unpublished arxiv profile agrees 
well with ours here.}  Again, we have little basis for proceeding with our interpretation 
of the B1953+29 profile unless its putative core component shows an adequate width 
reflecting that of the polar cap.  For this 6.1-msec pulsar, the intrinsic core width 
$2.45\degr P^{-1/2}$ [ET VIa, eq.(2)] is expected to be some 31.3\degr.  Fits to 
the central component discussed in the Appendix give a width of 34.5\degr\ with a 
fitting error of less than a degree.  The larger error here, though, is systematic:  a 
Gaussian function may not fit the core feature well and the adjacent overlapping 
components imply significant correlations.  The fitting then indicates a core width 
which is unlikely to be larger than the fitted value but possibly a little smaller. Thus 
there is strong indication that this is a core, and that the width reflects a magnetic 
colatitude---via the $\csc\alpha$ factor---close to orthogonal, that is, some 65\degr\ 
or more.  The leading feature (preceding the core by about 163\degr) would then 
support a nearly orthogonal geometry if indeed this is an interpulse---and we will 
make this assumption.\footnote{The unusual 360\degr\ total PPA rotation also 
suggests an inner (that is poleward) sightline traverse.}  

Now we can also see from Figure~\ref{fig6} that the PPA rate is about --3\degr/\degr.  
Following the quantitative geometric analysis of ET VI using the profile dimensions 
(see the Appendix and Table~A2), the results are given in Table~\ref{tab4}, 
where for a magnetic colatitude $\alpha$ of 65\degr\ the characteristic emission height 
would be only some 81 km (rather than the 110-120 km typical of inner-cone characteristic 
heights of normal pulsars).  An A/R analysis of the B1953+29 profile, assuming that 
the core component is at the longitude of the magnetic axis, results in the conclusion 
that the conal center precedes the core by 11$\pm$3\degr, and that in turn corresponds 
a relative core-cone emission height difference of 27$\pm$5 km within a light-cylinder 
radius of 293 km.

\subsection{J1022+1001}
PSR J1022+1001, discovered by Camilo \etal\ (1996), is a recycled millisecond pulsar 
with a rotation period of 16 msec.  It is in a binary system with an orbital periodicity of 
7.8 days.  The pulsar has been part of several long term timing programs, but is well 
known to time poorly\footnote{However, with adequate calibration the pulsar has 
recently been shown to time well (van Straten 2013)}.  It has been observed over a 
broad frequency band,  (\eg Dai \etal\ 2015; Noutsos \etal\ 2015), and below 1 GHz 
its profile is clearly comprised of three prominent components.  The underlying PPA 
traverse has a characteristic S-shaped RVM form with a ``kink'' under the central 
component similar that seen in B0329+54 by Mitra \etal\ (2007), such that the central 
slope is some 7\degr/\degr\ (Xilouris \etal\ 1998; Stairs \etal\ 1999).  Some studies 
argue that the pulsar exhibits profile changes on short timescales, an effect perhaps 
similar to mode changing in normal pulsars (see Kramer \etal\ 1999; Ramachandran 
\& Kramer 2003; Hotan \etal\ 2004; and Liu \etal\ 2015).  Our three pairs of Arecibo 
observations at 327 and 1400 MHz also suggest slightly different linear and circular 
polarization at different times, while confirming the basic profile structures seen in the 
above papers and the star's profiles shown here.  

In Fig.~\ref{figA5} we show Gaussian component fits to J1022+1001 using profiles from 
Stairs \etal\ (1999) (Jodrell Bank profiles downloadable from the European Pulsar 
Network  database\footnote{http://www.epta.eu.org/epndb/ }).  We find that these profiles 
can be decently fitted with three components, the central one having a width of 17\degr\ 
at 610 MHz and 14\degr\ at 410 MHz, roughly close to the polar cap size of 19\degr.  In 
these fits the location and width of the leading component are highly correlated and 
cannot be constrained very well, with some effect also on the central component.  This 
said, the effect of A/R in terms of the core center lagging the center of the overall profile 
is clearly seen in this pulsar.  The magnitude of this lag is about 6.5\degr\ at 410 MHz 
and 7\degr\ at 610 MHz, where we find the center of the overall profile based on the outer 
Gaussian-fitted component half-power points in Table~A2.  These estimates place the 
conal emission at about 20 km above the core emission. 

In a short communication Mitra \& Seiradakis (2004) used an A/R model to estimate the 
emission height across the pulse.  They argued that the pulsar could be interpreted 
in terms of a central core and a conal component pair as shown in their fig. 2.  They 
found a slight difference in emission height between the core and conal emission, 
which then had the effect due to A/R of inserting a kink in the PPA traverse similar to 
what is observed.  For this to happen they suggest that the core emission originates 
closer to the neutron star surface, and the conal emission region is then about 25 km 
above that of the core emission, similar to estimates obtained from the profile analysis 
mentioned above.





\section{Summary of Observational Results}
\label{sum}
MSP J0337+1715 apparently provides a rare opportunity to study a core/double-cone 
emission beam configuration in the recycled and relatively weak magnetic field system 
of a millisecond pulsar.  Slow pulsars with five-component profiles showing the core and 
both conal beams are unusual---because our sightline must pass close to the magnetic 
axis to encounter the core and resolve the inner cone---but such stars exhibit the most 
complex beamforms that pulsars normally seem capable of producing.  

The conal dimensions of J0337+1715 and A/R analysis provide a novel and consistent 
picture of the emission regions in this MSP.  The A/R analysis argues that the inner conal 
emission is generated at a height of some 70 km and the outer cone then at about 85 km, 
and adjustment of the quantitative geometry for the estimated $s_L$ values provides 
roughly compatible emission heights.  No evidence of conal spreading is seen over the 
nearly three octave band of observations, again suggesting that all the emission is 
produced within a narrow range of heights.

Questions have remained about whether MSPs radiate in beamforms similar to those of 
slower pulsars.  When putative core components have been seen in MSPs---as in the 
roughly five-componented profiles of J0437--4715---they are usually narrower than the 
polar-cap angular diameter and thus cannot be core emission features in the manner 
known from the slow pulsar population.  This might be because MSP magnetic dipoles 
are not centered in the star, which in turn might result from the field destruction during their 
accretional spinup phase (\eg Ruderman 1991).  Or conversely, perhaps in J0337+1715 
this field destruction was more orderly or less disruptive than in many other MSPs.  

Core/cone beam structure had heretofore been identified in a few MSPs---B1913+16, 
B1953+29 and J1022+1001---and not systematically assessed in these until recently 
because the available polarimetry made quantitative geometric modeling difficult and 
inconclusive.  Here we have been able to carry out such analyses that argue strongly 
for core/cone structure in these three other MSPs.  In each case we find core widths 
that are compatible with the full angular size of their (dipolar) polar caps at the surface.  
We further find characteristic emission heights for these pulsars that are all smaller or 
much smaller than expected for the inner cones of normal pulsars.  A/R analyses then 
provide more physical emission heights:  For B1913+16, a 59-ms MSP with the largest 
light-cylinder of 2820 km, its properties are most like those of slow pulsar inner cones.  
For each of the other three, however, with much faster rotations and smaller 
magnetospheres, both the quantitative geometry and A/R analyses indicate radio 
emission from lower altitudes of only a few stellar radii.  

These results can embolden us to recognize core/cone structure in other MSPs, classify 
them and study their emission geometry quantitatively in order to determine if, in fact, 
some few other core/cone profiles can be recognized among the many MSPs with 
inscrutable ones.  These results underscore that while the few MSPs here seem to 
exhibit such structure, many or most appear not to.  Thus these results may begin to 
provide a foundation for exploring the consequences of accretional magnetic field 
destruction during MSP spinup.  It may well be that orderly core/cone beam structure 
signals and depends on a nearly dipolar magnetic field configuration at the height of 
the radio-emission regions (Rankin \etal\ 2017).  Therefore, for reasons to be learned 
in the future, some MSPs may be able to retain a sufficiently dominant dipolar 
magnetic field in their radio emission regions despite its dramatic weakening by 
accretion---whereas, most other MSPs apparently do not.   

\begin{table*}
\caption[]{Summary of plasma properties for MSP radiation.}
\begin{center}
\begin{tabular}{lccccccccccc}
\hline
Pulsar   & P  & $\dot{P}$ & Beamform & $\nu_{obs}$ & h & $\nu_B$ &  $\nu_p$  & $\nu_{cr}$ & $R$ & R$_{lc}$  \\
         &(ms) &(10$^{-20}$ s/s)&                 & (GHz)& (km) & (GHz)   & (GHz)     & (GHz)      &          & (km)        \\
\hline
J0337+1715& 2.732 & 1.76    & Inner Cone & 1.4  & 69 & 19 &15$-$146 &11$-$87 & 0.53 & 130  \\
          &       &         & Outer Cone & 1.4  & 83 &  11 &11$-$111 &10$-$79 & 0.64  & 130  \\
          &       &         &            &      &    &     &         &         &      &      \\
B1913+16 & 59.0   & 86        & Inner Cone & 1.4 & 126 & 101  &7$-$73  &2$-$14 & 0.04 & 2817 \\
         &        &           &               &     &     &      &        &       &       &     \\
B1953+19 & 6.13    & 2.97      & Inner Cone & 1.4 &  27 & 614  &55$-$555&12$-$93& 0.09  & 293 \\
         &        &           &             &                    &     {\it 81}&{\it 23}&{\it 11$-$107}&{\it 7$-$54}&{\it 0.28}  & 293 \\
         &        &           &               &     &     &      &        &       &       &     \\
J1022+1001& 16.4  & 4.3       & Inner Cone & 1.4 &  25 &1530  &53$-$535& 7$-$59& 0.03 & 786  \\
         &        &           &             &                    &    {\it 45}&{\it 262}&{\it 22$-$221}&{\it 5$-$44}&{\it 0.06} & 786 \\
          \hline
\end{tabular}
\end{center}
{\it Notes:} The frequency $\nu_{B}$ is the cyclotron frequency estimated for $\gamma_s = 200$,
frequency $\nu_p$ estimated for $\gamma_s = 200$ and two values of $\kappa = 100, 10^4$, 
the cyclotron frequency estimated for two values of $\Gamma_s = 300, 600$.  Italicized values follow from the geometric model (characteristic) heights in Table~\ref{tab4}.
\label{tab5}
\end{table*}

\section{Physical Analysis of Emission-Region Conditions}
\label{phys}

Other recent A/R analyses of slower pulsars (\eg Mitra \etal\ 2016 and references 
therein) and theoretical developments (\eg Gil \etal\ 2004; Melikidze \etal\ 2015 and 
references therein) strongly argue that the radio emission in slower pulsars comes 
from altitudes of about 500 km.  By then estimating the magnetic field strength, the 
Goldreich-Julian (1968) plasma density $n_{GJ} = {\bf\Omega} \cdot {\bf B}/2 \pi c$ 
(rotational frequency $\Omega = 2 \pi/P$, $B$ is the magnetic field, and $c$ the 
lightspeed), and the plasma, cyclotron and curvature frequencies at these heights 
of emission, we are able to show that the radiation emerges from regions where the 
three latter frequencies are all much greater than the frequency of the emitted 
radiation.  This indicates that the mechanism of radiation must be the curvature 
process and the emitting entities charged solitons (\eg Mitra, Gil \& Melikidze 2009; 
Mitra, Rankin \& Arjunwadkar 2016; Mitra \& Rankin 2017), which in turn connects 
with constraints on the physical propagation modes (ordinary, extraordinary) of the 
observed radiation (Rankin 2015).  

We then here assess the results of a similar modeling of the physical properties of 
the emission regions in the four MSPs studied above.  A summary of the estimated 
plasma properties for each millisecond pulsar at its A/R-determined emission height 
is given in Table~\ref{tab5}.\footnote{Values are also given in italics for B1953+29 
and J1022+1001 corresponding to the geometric model (characteristic) heights in 
Table~\ref{tab4} where the A/R height may only measure the difference between the 
core and conal emission heights.}  The cyclotron frequency $\nu_{B}$ is estimated for a 
secondary plasma Lorentz factor $\gamma_s$ of 200.  The plasma frequency $\nu_p$ 
is estimated for $\gamma_s$=200 and two values of the overdensity $\kappa$=$n_s/n_{GJ}$ of 
100 and 10$^4$, and the curvature frequency $\nu_{cr}$ for two values of $\Gamma_s 
= 300, 600$, the Lorentz factor of the emitting soliton.  The radius of the light cylinder 
is given by  $R_{lc} = c P/2\pi$, and the model height of the radio emission $r$ in 
terms of light cylinder distance is given by $R = r/R_{lc}$.  

The Table~\ref{tab5} values show that the radio frequency is well less than conservative 
estimates of the cyclotron, plasma and curvature frequencies in every case.  This in turn 
demonstrates not only that MSP emission functions physically in a manner that is very 
similar or identical to slower pulsars, but it also indicates that the mechanism of radiation 
must be the curvature process and the emitting entities charged solitons as in slower 
pulsars (Mitra, Rankin \& Arjunwadkar 2016).  Very similar conclusions can be reached 
via the conclusion that the 22.7-ms pulsar J0737--3039A has an emission height of some 
9-15 stellar radii (Perera \etal\ 2014).

\section{Conclusions}
\label{con}
MSP J0337+1715 together with the other three pulsars studied above apparently 
show us that some MSPs are capable of core/(double)-cone emission---and thus 
that the engines of emission in MSPs are not essentially different than those in 
slower pulsars.  In particular, we have been able to estimate the emission heights 
of the four MSPs using both geometric modeling and A/R, and all indicate that 
the emission regions lie deep within the larger polar flux tubes of their shallower 
magnetospheres.  We also find as for the slower pulsars that in their radio emission 
regions the cyclotron, plasma and curvature-radiation frequencies are much higher 
than those of the emitted radiation, strongly arguing for the curvature emission 
process by charged solitons.

\section*{Acknowledgments}   We thank Paul Demorest for assistance in analyzing 
a Green Bank Telescope observation of B1913+16 and Michael Kramer for use of 
his bfit code.  Much of the work was made possible by support from the US 
National Science Foundation grant 09-68296 as well as NASA Space Grants.  
One of us (JMR) also thanks the Anton Pannekoek Astronomical Institute of the 
University of Amsterdam for their NWO Vidi/Aspasia grant (PI Watts) support.  
Another (JvL) thanks the European Research Council for funding under Grant 
Agreement n. 617199.  JMW was supported by NSF Grant AST 1312843 
Arecibo Observatory is operated by SRI International under a cooperative 
agreement with the US National Science Foundation, and in alliance with SRI, 
the Ana G. M\'endez-Universidad Metropolitana, and the Universities Space 
Research Association.  This work made use of the NASA ADS astronomical 
data system.

{}

\appendix
\setcounter{figure}{0} 
\renewcommand{\thefigure}{A\arabic{figure}}
\setcounter{table}{0} 
\renewcommand{\thetable}{A\arabic{table}}
\setcounter{footnote}{0} 
\renewcommand{\thefootnote}{A\arabic{footnote}}
\section*{APPENDIX: Results of  Profile Modeling}
\label{model}

\subsection{J0337+1715}
In order to intercompare our analyses in the four bands conveniently, they 
should be well aligned in time.  An attempt was made to align the profiles 
according to their timing and dispersion delay, which was roughly satisfactory 
(apart from the 2.0-GHz band which was found to be early by about 10\degr\ 
and the 430-MHz one that was late by about 5\degr.  The adjustments brought 
the centers of the putative ``caustic'' features into close alignment as can be seen 
in the models of Table~A1.   

Figure~\ref{figA1} shows the 1391-MHz and other profiles with the various 
model components.  Similar modeling efforts were carried out for each of the 
profiles, and the results are tabulated in Table~A1.  The inner and outer 
cones and core are modeled with Gaussian components, and one further Gaussian 
is used to model the putative caustic component on the trailing edge of the trailing 
inner conal component.  The leading conal components then align within an 
{\it rms} of about a degree, and the trailing ones about twice that.  Clearly, the 
Gaussian functions model the profile components only poorly.  Most components 
have complex and asymmetric shapes, and the fitting results seem adequate only 
to estimate the widths and centers of the five components approximately. 

However, it is the center of the core component which in principle falls close to 
the magnetic axis, but this point can only be estimated by modeling due to the 
conflation of the core with the trailing components.  In particular, the Gaussian 
model for the core component's shape at 1391-MHz is clearly poor.  The 
modeling efforts do show that the space is adequate before and under the trailing 
components for a core of the expected 67.5\degr\ halfwidth.  Recent evidence 
indicates that some cores may have an unresolved double form, and a model 
using two Gaussians of half the expected width and separated by this halfwidth 
does model the 1391-MHz feature somewhat better (dashed magenta curve).  

\begin{figure}
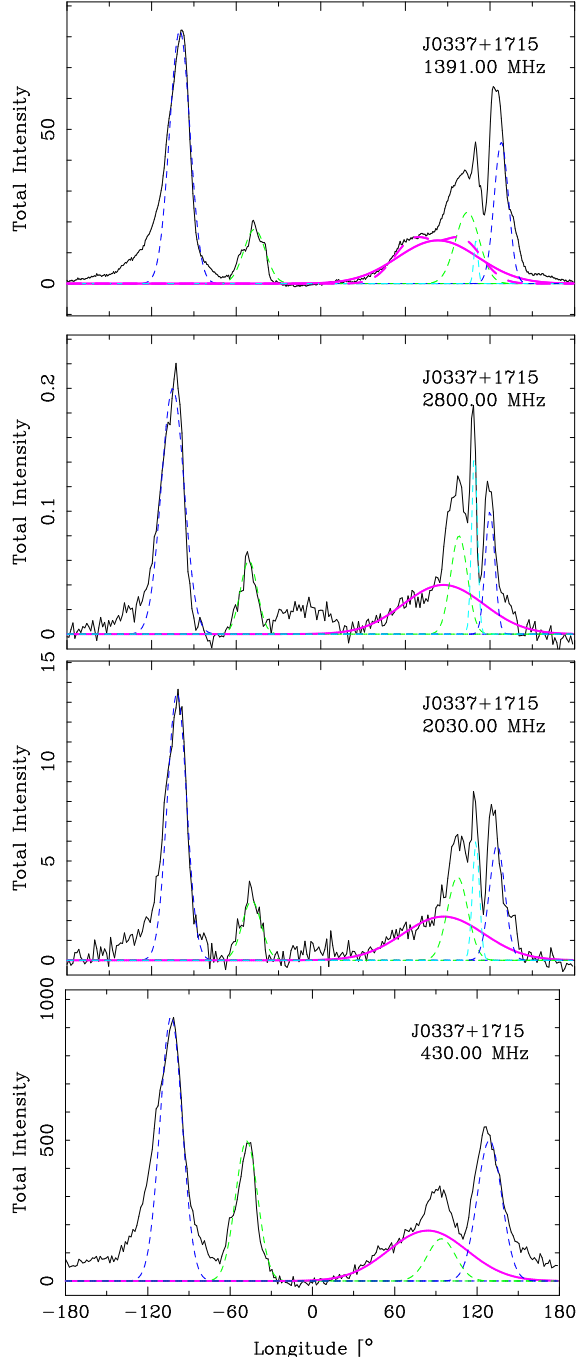

\begin{center}
\mbox{\includegraphics[height=75mm,angle=-90.]{sJ0337+1715l.uvm_gauss.ps}}
\vskip 2mm
\mbox{\includegraphics[height=75mm,angle=-90.]{sJ0337+1715su_gauss.ps}}
\mbox{\includegraphics[height=75mm,angle=-90.]{sJ0337+1715sl_gauss.ps}}
\mbox{\includegraphics[height=75mm,angle=-90.]{sJ0337+1715u_gauss.ps}}
\caption{Total power profiles of J0337+1715 as in Figs.~\ref{fig1} and \ref{fig3} 
showing the several Gaussian functions used to model the profiles and estimate 
the centers and dimensions of the components in Tables~A1, \ref{tab2} and 
\ref{tab3}.  The outer conal Gaussian functions are shown in dashed blue, 
the inner dashed green, and the core solid magenta---and the caustic feature 
dashed cyan.  An unresolved double model is also shown for the 1391-MHz 
profile (dashed magenta); see text.  The model dimensions are given in Table~A1.}
\label{figA1}
\end{center}
\end{figure}

\begin{table}
\begin{center}
\caption{J0337+1715 Component Modeling Summary.}
 \begin{tabular}{lcccccc}
 \hline
 \hline
	&LOC&LIC&Core&TIC&TOC&``CC'' \\
 \hline
 \multicolumn{6}{c}{\bf 1.4-GHz Band} \\
 C (\degr) & --100 & --47 & 83 & 99 &128 &110  \\
 W( \degr) &  17  &  18   & 67.5 &   20  & 12 &   3.5   \\
 A  &  82.0  &  17.6   & 14.0 &   23.1  & 45.7 & 14.0  \\
 \multicolumn{6}{c}{\bf 2.8-GHz Band} \\
 C (\degr)& --105  &--51 & 87  &   98  & 124  & 109  \\
 W (\degr)&  19  &  14   & 67.5  &  14   &  8 &    4.7  \\
 A & 0.20   &  0.06 & 0.04 &  0.08 & 0.10 & 0.15 \\
 \multicolumn{6}{c}{\bf 2.0-GHz Band} \\
 C (\degr) &--102 &--49  & 87  & 97 & 125 &110  \\
 W (\degr)&   16   &  16  & 67.5  &  17  &   13  &     5.8  \\
  A &   13.5 &    3.0 &   2.2 &     4.2 &     5.8 &     6.0 \\
 \multicolumn{6}{c}{\bf 327-MHz Band} \\
 C (\degr)&--103 &--48 & 84  & 94  &129  & ---   \\
 W (\degr) &   19  & 19 &  67.5  & 24 &  21 & --- \\
  A &    940 &  498 &  179 &  150 &   500 & --- \\
\hline
\end{tabular}
\end{center}
Notes: ``LOC/LIC'' Leading Inner/Outer Cones; ``TIC/TOC'' Trailing Inner/ Outer 
Cones; and ``CC'' Caustic Component.  ``C'' center; ``W'' half-power width; 
and ``A'' amplitude.
\label{tabA1}
\end{table}

\begin{table}
\begin{center}
\caption{Component Fitting Summary for pulsars B1913+16, B1953+29 and J1022+1001.}
 \begin{tabular}{lccccc}
 \hline
 \hline
Pulsar && LC&Core&TC& IP \\
(GHz) &&(\degr)&(\degr)&(\degr) &(\degr) \\
 \hline
 B1913+16&C & --23.8 &  --2.8  & 15.8  & ---  \\
          	& W &  13.6  &  16.6   & 15.0 & ---  \\
          	& A  &  0.06  &  0.03   & 0.05 & ---  \\
 B1953+29&C & --52.2 &  --2.8  & 15.5  & --165.6  \\
          	& W &  13.6  &  16.6   & 15.0 & 25.9  \\
          	& A  &  132e3  &  55e3   & 81e3 & 9e3  \\
 J1022+1001\\
    410 MHz & C &  --12.5  &  0.4   & 12.5 & --- \\
		& W   &   27.9  &  13.6   & 6.0 & --- \\
		& A    &  0.34  &  0.77   & 0.73 & --- \\
    610 MHz & C &  --15.3  &  3.4   & 15.1 & --- \\
		& W  &  18.9  &  17.8   & 5.4 & --- \\
		& A  &  0.15  &  0.59   & 0.78 & ---  \\
 \hline
\end{tabular}
\end{center}
\label{tabA2}
\end{table}

\begin{figure}
\begin{center}
\mbox{\includegraphics[height=75mm,angle=-90.]{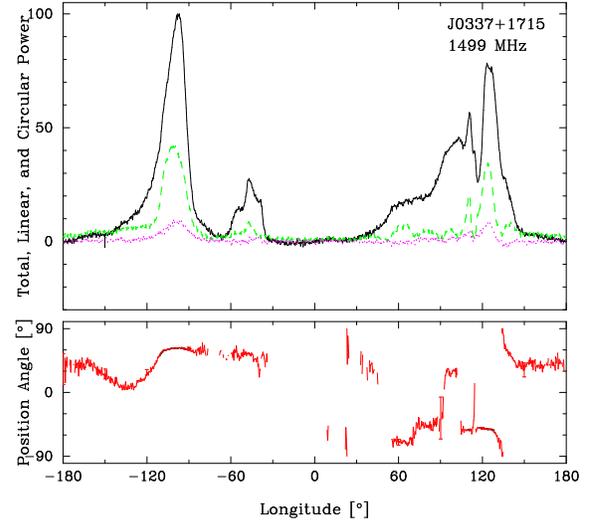}}
\caption{Total 1.4-GHz Green Bank Telescope GUPPI profile of pulsar J0337+1715 as 
in Fig.~\ref{fig1}.  The profiles compare well with each other despite concerns that the 
polarization might be distorted by reflections for a source that must be tracked at 
Arecibo fully under its triangular supporting structure.  The Green Bank Telescope is 
designed to avoid all such issues.  Arecibo's PUPPI and the latter's GUPPI are almost 
identical instruments.}
\label{figA2}
\end{center}
\end{figure}

\begin{figure}
\begin{center}
\mbox{\includegraphics[width=63mm,angle=-90.]{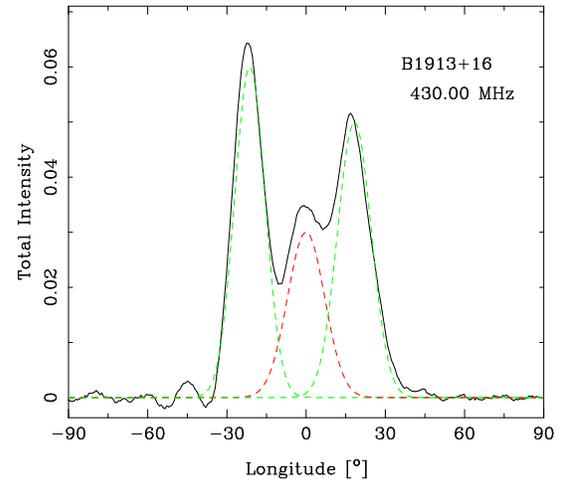}}
\caption{Arecibo 430-MHz profile of B1913+16 from 1990 August 18 showing the 
three Gaussian functions used to fit the profile and estimate the centers and 
dimensions of the components in Tables~A2 and \ref{tab4}.  The conal 
Gaussian functions are shown in green, and the central core in red.  The 
baseline ripples on the leading edge may result from interference.}
\label{figA3}
\end{center}
\end{figure}

The Gaussian component modeling of the estimated core centers gives  
reasonably consistent results for the four bands as seen again in Fig.~\ref{figA1} 
and Table~A1.  In each case the core width was fixed at its expected 
67.5\degr\ value, and its amplitude was also adjusted to be about 15\% of the 
two leading conal component's amplitudes in order to control the correlation 
of its amplitude with that of the trailing inner conal component.  Within these 
assumptions, the estimated core center position falls at +85\degr$\pm$2\degr\ 
longitude.  However, the probable errors are much larger.  The A/R analysis 
above uses the differences between the conal component centers and the 
core center.  We have seen that the model differences across the bands of a 
conal component center are only one or two degrees; however, a different 
modeling assumption might shift the core center by a larger amount.  We 
therefore propose to adopt a core center value of +85\degr$\pm$10\degr\ 
to accommodate these various uncertainties.

\subsection{B1913+16}
The Binary Pulsar B1913+16 can no longer be observed to clearly show its 
three components and putative core-cone structure at frequencies around 
430 MHz.  Its secularly changing profiles show that its emission beam is 
precessing, shifting the orientation of our sightline's traverse through it, such 
that eventually its beam will no longer point in our direction (Weisberg \& Taylor 
2002).  Therefore, we make use here of both archival and current observations.  
Please refer to the well measured Arecibo profiles of Blaskiewicz \etal\ (1991; 
their fig. 17); Figure~\ref{fig5} gives a current Arecibo 1390-MHz profile; and 
we use another archival Arecibo observation in Figure~\ref{figA2} to determine 
the centers and widths of its three profile components by Gaussian fitting.  

The fitting results are given in Table~A2.  Here the Gaussian fits worked 
well as can be seen in the figure, implying that here the three components can 
be closely modeled using Gaussian functions.  These dimensions are used above 
to assess the half-power width of the central components so as to vet it as a core 
component, and in Table~\ref{tab4} to quantify the geometry of the pulsar's 
cone after Paper ET VI.

\subsection{B1953+29}
\begin{figure}
\begin{center}
\mbox{\includegraphics[width=63mm,angle=-90.]{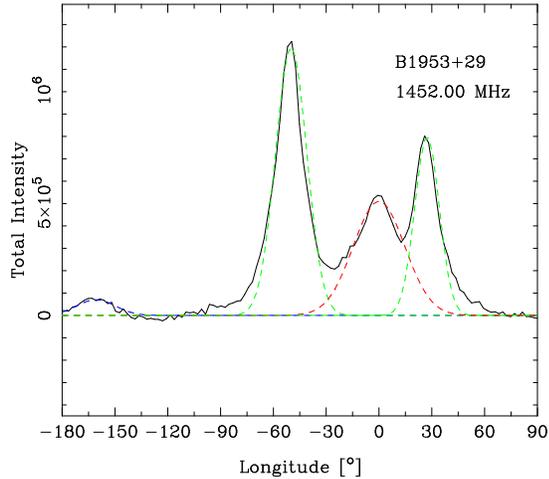}}
\caption{Total power profile of B1953+29 at 1452 MHz as in Fig.~\ref{fig5}, 
showing the several Gaussian functions used to fit the profile and estimate 
the centers and dimensions of the components in Tables~A2 and 
\ref{tab4}.  The conal Gaussian functions are shown in green, and the central 
core in red.}
\label{figA4}
\end{center}
\end{figure}

Similar Gaussian fitting was carried out for MSP B1953+29 using the 1452-MHz 
Arecibo observation in Fig.~\ref{fig5} above.  The results of the fits for the three 
profile components and possible interpulse are shown in Figure~\ref{figA4} where 
it is clear that the modeling was again in this case successful.  Gaussian functions 
fit all four profile features well, and their fitted centers and widths are again given 
in Table~A2.  The results are also used above to assess whether the core 
width is compatible with a core identification and to provide an analysis of the conal 
beam geometry in Table~\ref{tab4}.

\subsection{J1022+1001}
\begin{figure}
\begin{center}
\mbox{\includegraphics[height=74mm,angle=-90.]{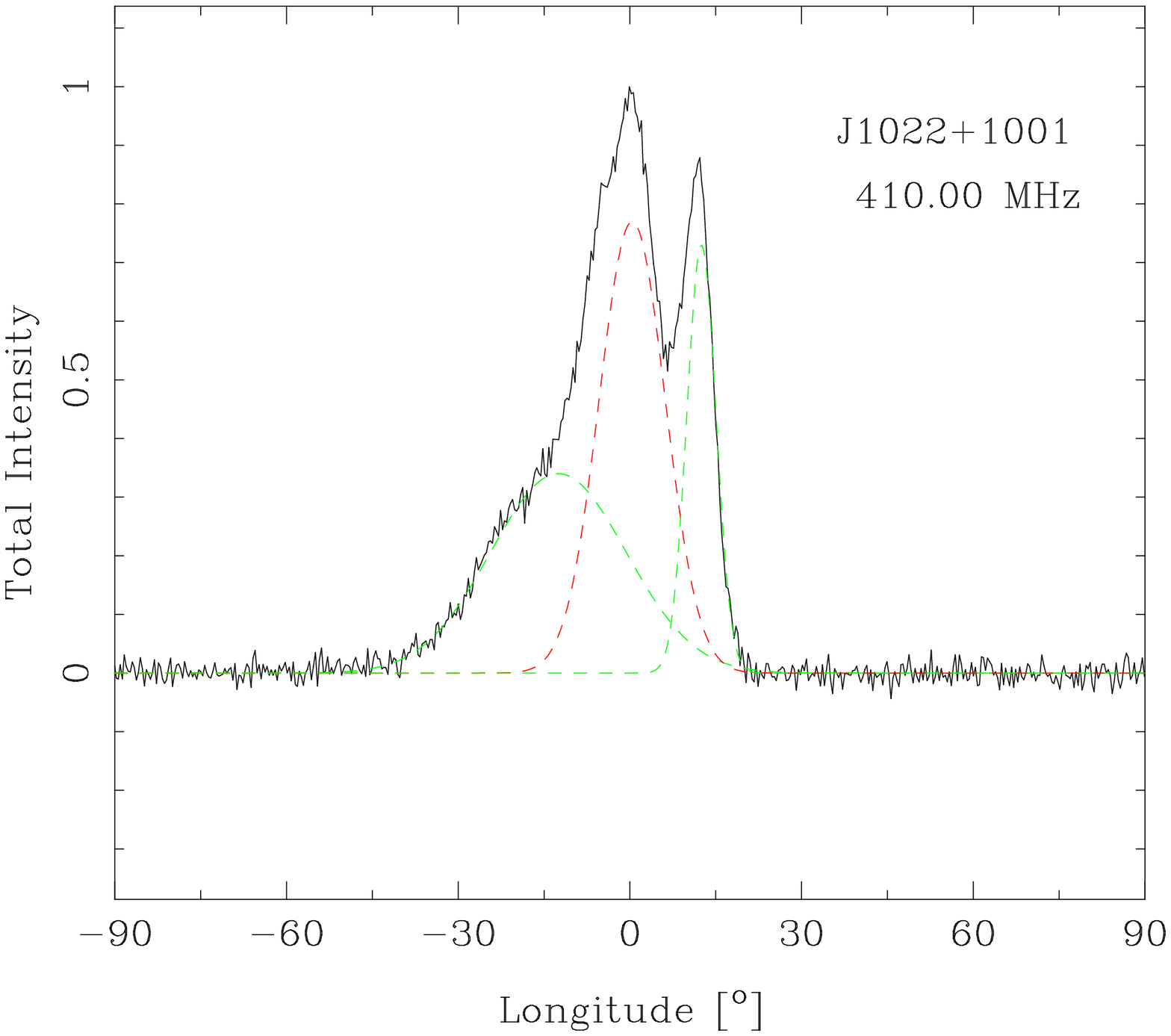}}
\mbox{\includegraphics[height=74mm,angle=-90.]{J1022+1001_610_epn_asc_gauss.ps}}
\caption{Total power profiles of J1022+1001 at 410 MHz (upper panel) and 
610 MHz (lower panel) from Jodrell Bank Observatory, showing the several 
Gaussian functions used to fit the profile and estimate the centers and dimensions 
of the components in Tables~A2 and \ref{tab4}.  The conal Gaussian 
functions are shown in green, and the central core in red.}
\label{figA5}
\end{center}
\end{figure}

We also studied the profiles of pulsar J1022+1001.  The results of the two fits for the 
pulsar's three profile components are given in Figure~\ref{figA5}.  Gaussian functions 
fit the three profile features well, and their fitted centers and widths are again given 
in Table~A2.  Kramer \etal\ (1999) fitted the overall pulsar profile at 1.4 GHz 
with five Gaussian components, and the star's profile does exhibit a narrow ``cap''-like 
feature at 1 GHz and above somewhat similar to J0337+1715's ``caustic'' one.  
Indeed, our observations show that the relative intensity of this narrow ``cap'' feature 
increases very rapidly with frequency from being undetectable at meter wavelengths, 
just so at 1 GHz and dominant at 1.7 GHz.  The alignment of the meter and centimeter 
profiles is very nicely shown in the recent paper by Noutsos \etal\ (2016; fig. 11).  We 
therefore conclude that its behavior is unlike anything previously seen in cores and 
can be ignored for the present purposes.  The results are also used above to assess 
whether the core width is compatible with a core identification and to provide a 
quantitative analysis of the conal beam geometry in Table~\ref{tab4}.

\end{document}